\documentclass[epj]{svjour}
\usepackage{graphics}
\begin{document}
\title{Estimate of the charming penguin contributions to $B \to \pi \pi$}
\author{Bla\v zenka Meli\'c\inst{1}
}                     
%
%
\institute{Rudjer Bo\v skovi\'c Institute, Theoretical Physics Division, HR-10002 Zagreb, Croatia}
\date{Received: date / Revised version: date}
%
\abstract{
We consider the problem of factorization in $B$ decays
and illustrate the calculation of nonfactorizable contributions
employing the
QCD light-cone sum rule method. We present a more detailed 
calculation of
the ``charming penguin'' contributions
as a potential source of the substantial nonfactorizable $O(1/m_b)$ effects
in the $B \rightarrow \pi\pi$ decay.
Although the
predicted corrections are not sizable, by calculating the $CP$ asymmetry we illustrate how such
corrections can
accumulate to
a visible effect. In conclusion,
nonfactorizable contributions in
nonleptonic $B$ decays into charmonium are briefly discussed.
} 
\maketitle
\section{Nonleptonic $B$ decays and factorization}
                                                                                         
For a long time the naive factorization method, in which the matrix
element of the four-quark operator is approximated by the product of two matrix
elements of the bilinear quark currents, 
was considered as a sufficiently precise tool for
estimating matrix elements emerging in the amplitude of nonleptonic weak $B$ decays.
Nowadays, in order to
make real use of the already very precise experimental data, we are forced to provide 
a more accurate estimate of nonleptonic decays, in particular of the nonperturbative
part of the decay amplitude.
Therefore, the question about the applicability of the factorization and
the size of
nonfactorizable corrections naturally emerged. The question was particularly raised in the work
\cite{Rome}, where it was argued that in the charmless $B$ decays there 
could exist large ${\cal O}(\Lambda_{QCD}/m_b)$ corrections
and large strong phases coming from the "charming penguins".

There are several models which one can apply for the calculation of matrix elements of
$B$-meson weak decays beyond
the naive factorization \cite{Sanda,BBNS,AK}.
By using the QCD factorization approach \cite{BBNS} 
one can show that the exclusive $B$-decay amplitude
in  the $m_b \to\infty$ limit can be expressed in terms of the factorizable part
and the calculable $O(\alpha_s)$ nonfactorizable correction.
However, because of the arguments given above,
the nonfactorizable subleading effects in the decay amplitude, 
suppressed by inverse powers of $m_b$,
could be  important and have to be investigated.
Estimates of nonfactorizable
contributions in $B$ decays, including the power-suppressed $O(1/m_b)$ 
contributions, can be obtained \cite{AK}
using the method of QCD light-cone sum rules (LCSR). 
In particular, nonfactorizable contributions to nonleptonic decays 
such as $B \to \pi \pi $ and $B \to \pi K$ become
interesting for a more precise constraint on the $\gamma = arg (V_{ub})$ angle of the CKM
matrix. In connection with this problem, it is worth mentioning that there are also several
strategies used to determine the $\gamma$ angle from
$B \to \pi \pi $ and $B \to \pi K$ based on
the isospin and SU(3) relations. Unfortunately, the theoretical accuracy of these relations is
limited and it has to be
improved by calculating the SU(3) breaking effects, which can also be addressed by  the LCSR
method \cite{Khodjaproc}.
                                                                                         
\section{Matrix elements for $B \to \pi \pi $ from LCSR}
                                                                                         
The LCSR expression for the $B\to \pi\pi$ hadronic
matrix element of the ${\cal {O}}_i$ operator of the weak
Hamiltonian is derived
by the procedure presented in detail in
\cite{AK,KMU}. One starts by introducing the correlation function
\begin{eqnarray}
F_{\alpha}^{({{\cal O}}_i)} &=&
i^2 \int \!d^4 x \,e^{-i(p-q)x}\!\int d^4 y \,e^{i(p-k)y}
 \nonumber \\
& & \times \langle 0 |
T \{ j_{\alpha 5}^{(\pi)}(y) {{\cal O}}_i(0) j_5^{(B)}(x) \} | \pi^-(q)
\rangle
\nonumber
\\
& =& (p-k)_{\alpha} F^{({{\cal O}}_i)}(s_1,s_2,P^2) + ...\,,
\label{eq:corr0}
\end{eqnarray}
where $ j_{\alpha 5}^{(\pi)} = \overline{u} \gamma_{\alpha} \gamma_5 d
$ and $ j_5^{(B)} = i m_b \overline{b} \gamma_5 d $
are  the quark currents
interpolating the pion and the $B$ meson, respectively.
By employing the dispersion relation technique and by assuming the quark-hadron
duality, we can write
the LCSR expression  for the hadronic matrix element
$
A^{({\cal O}_i)}( \bar{B}^0_d \to \pi^+ \pi^-)
\equiv \langle \pi^-(p)\pi^+(-q) |
{\cal O}_i |\bar{B}^0_d(p-q) \rangle
$ as \cite{AK}
\begin{eqnarray}
f_\pi f_B A^{({\cal O}_i)}( \bar{B}^0_d \to \pi^+ \pi^-)
e^{-m_B^2/M_2^2}
&=&
\nonumber \\
& & \hspace*{-5.5cm} 
\int\limits_{m_b^2}^{s_0^B} \!ds_2
e^{-s_2/M_2^2}\!
\Bigg \{
\!\int\limits_0^{s_0^\pi} ds_1 e^{-s_1/M_1^2}
\mbox{Im}_{s_2}\mbox{Im}_{s_1} F^{({{\cal O}}_i)}(s_1,s_2,m_B^2)
\Bigg\} \, , 
\nonumber \\
& & 
\label{eq:sumrule}
\end{eqnarray}
where $M_{1}$ and $M_2$ are the Borel parameters in the pion and
$B$-meson channels, respectively.
The parameter $s_0^\pi$ $(s_0^B)$ is the effective threshold parameter of the
perturbative continuum in the pion ($B$-meson) channel.
In the sum rule (\ref{eq:sumrule}) the
finite $m_b$ corrections are taken into account,
but numerically very small corrections of order $s_0^\pi/m_B^2$ are neglected.
                                                                                         
\begin{figure}[hb]
\begin{center}
\resizebox{0.50\textwidth}{!}{%
\includegraphics{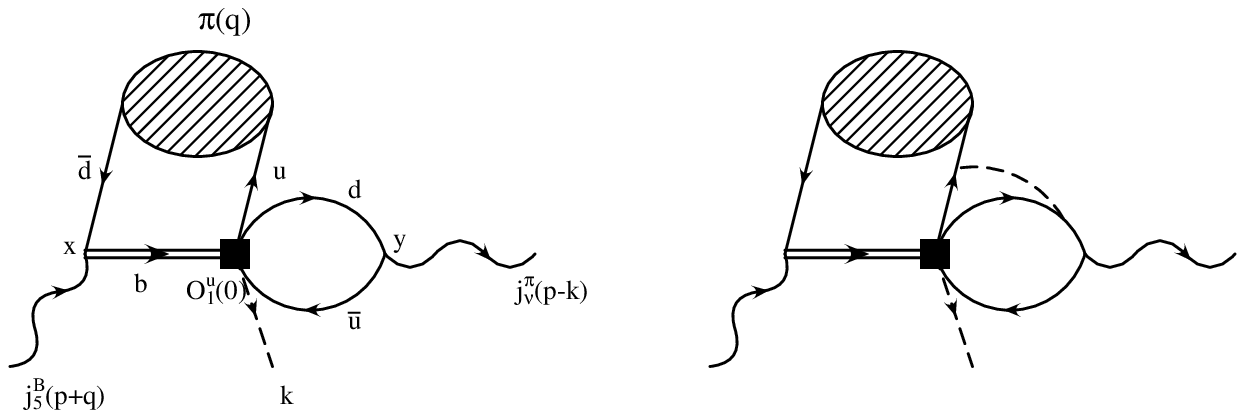}
}
\end{center}
\hspace*{1.7cm} (a)  \hspace*{4cm}(b) 
\begin{center}
\resizebox{0.25\textwidth}{!}{%
\includegraphics{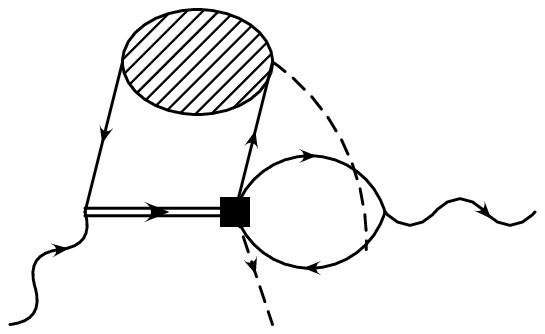}
}
\end{center}
\hspace*{4.0cm} (c)
\caption{Contributions in the emission topology:
(a) factorizable contribution from the ${\cal O}_1^u$ operator;
(b) an example of the $O(\alpha_s)$ correction;
(c) an example of the nonfactorizable diagram from the
${\tilde{\cal O}}_1^u$ operator.}
\label{fig:hard}
\end{figure}
                                                                                         
\subsection{Emission topology}
                                                                                         
The leading contributions in the emission topology are shown in Fig.1. The factorizable
part (Fig.1a) stems from the
contribution of the leading operator ${\cal O}_1^u = (\overline{d}\Gamma_{\mu}u)(\overline{u}
\Gamma^{\mu} b)$, mainly reproducing the naive factorization 
result in terms of the light-cone calculated $f_{\pi}$ and $F_{B \to \pi}$ form factor.
The hard corrections (Fig.1b) have not yet been addressed in LCSR, but
the corresponding contributions calculated in the $m_b \to \infty $ limit 
in QCD factorization \cite{BBNS} appeared to be small.
The soft nonfactorizable contribution due to the $\tilde{{\cal O}}_1^u$ operator (which is the
color-octet part of ${\cal O}_2^u = 1/N_c {\cal O}_1^u + 2 \tilde{\cal O}_1^u$ 
and expresses the exchange of a soft gluon between two
pions (Fig.1c)) was shown \cite{AK} to be
small, although of the
same size as the ${\cal O}(\alpha_s)$ correction mentioned above.
                                                                                         
\subsection{Penguin topology}
                                                                                         
Types of the
dominant penguin diagrams are shown in Fig.2 \cite{KMM}.
The main effect which we calculate arises from the $c$-quark loop annihilation into
a hard gluon (Fig.2a). In addition, there is the
quark-condensate contribution (Fig.2b) which after a  detailed analysis appeared to 
be a natural upper limit to all neglected contributions of multiparticle distribution 
amplitudes (DA's). 
The effects 
of the soft (low-virtuality) gluons
coupled to the $c$-quark loop \cite{KMM}
in the sum rule approach
manifest as multiparticle DA's and are therefore suppressed at least by ${\cal O}(\alpha_s/m_b)$.
Therefore, the nonfactorizable ${\cal O}(1/m_b)$ corrections from penguin loops are 
mainly of perturbative origin, and
both contributions from Fig.2. generate the  strong
rescattering phases in $B \to \pi \pi $ perturbatively by the well-known
BSS mechanism \cite{BSS}.
                                                                                         
\begin{figure}
\begin{center}
\resizebox{0.40\textwidth}{!}{%
\includegraphics{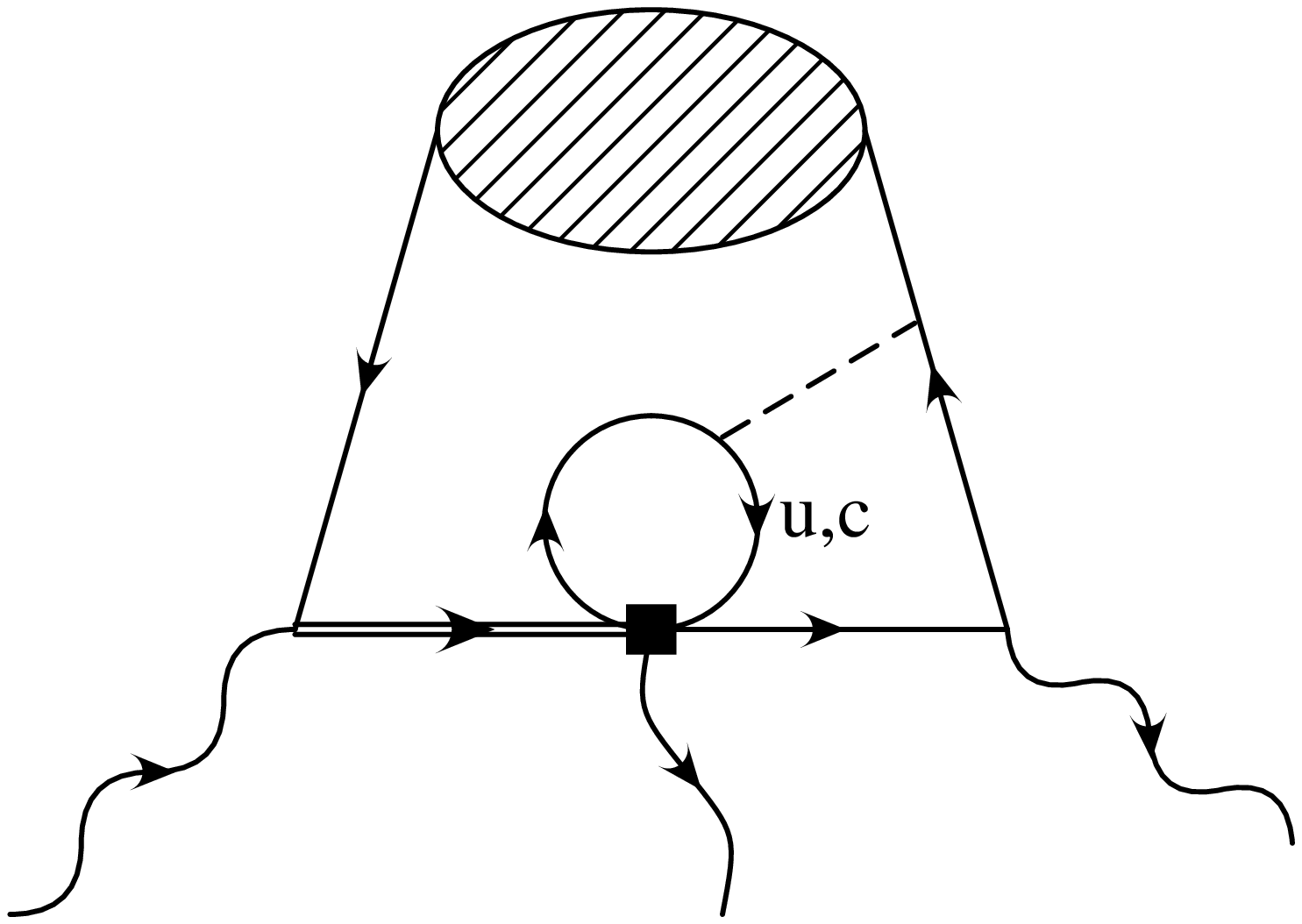}
\vspace*{0.5cm}
\includegraphics{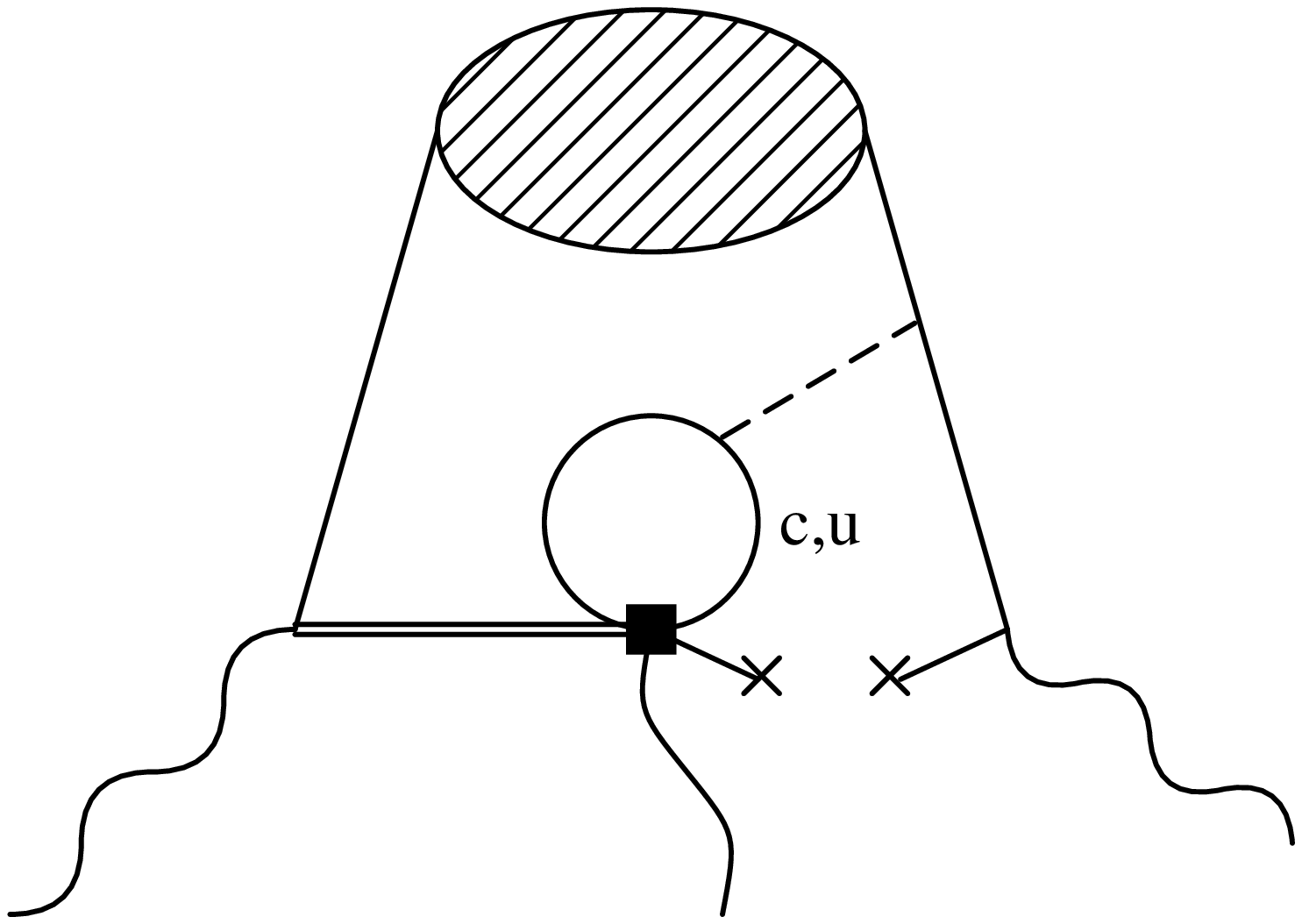}
}
\end{center}
\hspace*{2.5cm} (a)  \hspace*{3.0cm}(b)
\caption{Contributions in the penguin topology: (a) an example of the ${\cal O}(\alpha_s)$
nonfactorizable penguin amplitude; (b) an example of the chirally enhanced twist-3 contribution
to the penguin amplitude. The square stands for ${\cal O}_2^{u,c}$ and 
${\cal O}_{1-6}$ operators. The leading contribution from the ${\cal O}_8$ operator 
proceeds without the quark loop \protect{\cite{KMU,KMM}}.}
\label{fig:cond}
\end{figure}

\section{CP asymmetry in $\overline{B}^0_d \to \pi^+ \pi^-$}
                                                                                         
Penguin contributions  appear to produce
a notable effect in the direct CP asymmetry and we take the CP asymmetry
as a testing ground for the influence of the
$1/m_b$ corrections in the charming
penguin contributions.
Following \cite{KMM},
we concentrate on the direct CP asymmetry in the $B_d^0 \to \pi^+\pi^-$ decay, which is
given as
\begin{eqnarray}
& & a_{\rm CP}^{\rm dir}
\equiv (1-\left\vert \xi \right\vert^2)/(1+\left\vert \xi \right\vert^2) \, , 
\end{eqnarray}
where
$\xi = e^{-2i(\beta+\gamma)}(1 + R\,  e^{i\gamma})/(1 + R \, e^{-i\gamma})$
and $R \equiv -P/(R_bT)$.
Here $T$ is the contribution to the $B\to \pi\pi$ amplitude proportional to
$V_{ub} V_{ud}^* = |V_{ub} V_{ud}^*| e^{-i\gamma} $. It contains the
tree amplitude, the penguin-loop contractions of the current-current operators 
${\cal O}_{1,2}^u$, and
also the $V_{ub} V_{ud}^*$  proportional penguin ${\cal O}_{3-6}$ operator contractions.
The remaining contributions, being proportional to $V_{cb} V_{cd}^*$, are included in $P$.
The penguin-loop contractions of the current-current operators ${\cal O}_{1,2}^c$ represent the
main contribution to this part.
The factor $R_b = |V_{ub}||V_{ud}|/(|V_{cb}||V_{cd}|)$ is the
ratio of the CKM matrix elements.
                                                                                         
Both $T$ and $P$ amplitudes have strong phases; therefore, we have
$T = |T| e^{i\delta_T}$ and $P = |P| e^{i\delta_P}$  and the CP
asymmetry for $B_d^0 \to \pi^+\pi$ can be written as
\begin{eqnarray}
& & a_{\rm CP}^{\rm dir} =
\frac{-2 |R|\,  \sin(\delta_P - \delta_T)\, \sin \gamma}
{1 - 2 |R| \, \cos(\delta_P - \delta_T)\, \cos \gamma + |R|^2}\,.
\label{eq:acp}
\end{eqnarray}
All contributions shown in Figs.1 and 2 are calculated in LCSR at finite $m_b$. We 
also include 
the LCSR result for the gluonic penguin
contribution of the dipole operator ${\cal O}_{8g}$ \cite{KMU}. 
The electroweak penguin
contributions to $B \to \pi \pi$ are color-suppressed and negligible.
                                                                                         
The hard $O(\alpha_s)$ corrections to $T$ and $P$ amplitudes are known in
the $m_b \to \infty $ limit from QCD factorization \cite{BBNS}. We have
examined the influence of these contributions to the phases $\delta_T$ and $\delta_P$.
It appears that they are highly suppressed in comparison with the phases emerging from 
the penguin-loop contributions. Therefore, we have neglected $O(\alpha_s)$ corrections in
(\ref{eq:acp}).

\begin{figure}
\resizebox{0.45\textwidth}{!}{%
\includegraphics{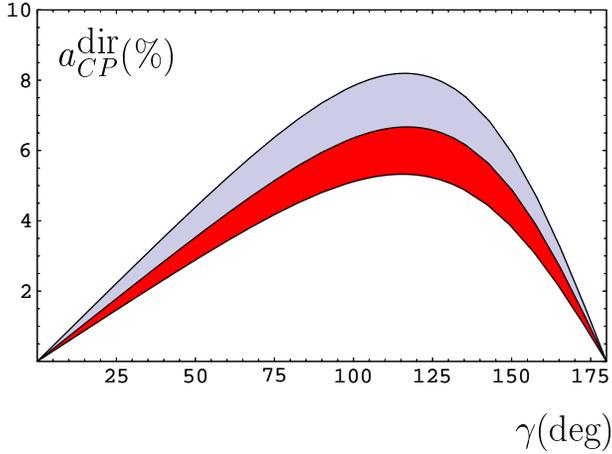}
}
\caption{Direct CP asymmetry in $B_d^0 \to \pi^+\pi^-$ as a function of the CKM
angle $\gamma$. The uppermost curve is the result obtained for
$m_b \to \infty$. The dark region is the LCSR result, with all uncertainties from the method included
(uncertainties in the CKM matrix elements are not taken into
account). The light region  shows the deviation from the $m_b \to \infty$
limit result.}
\label{fig:CP}
\end{figure}

In Fig.~\ref{fig:CP} we show $a_{\rm CP}^{\rm dir}$ as a function of
$\gamma$, calculated by
using the penguin
contributions estimated from LCSR at finite $m_b$ (dark region) and compare
the result in the infinite-mass limit that agrees with
the QCD factorization prediction \cite{BBNS} (the uppermost curve).
Both results are taken at the same scale $\mu_b\sim m_b/2$ as used in LCSR.
%
                                                                                         
The prediction shown in Fig.3 is not final,
annihilation effects are missing and the uncertainty in the CKM matrix elements is 
not taken into account either. However,
the figure nicely illustrates the size of $O(1/m_b)$ corrections  and
the difference between the results obtained at the finite $m_b$ and
in the $m_b\to \infty$ limit (the light region in Fig.3).
                                                                                         
\section{Conclusion}
                                                                                         
We have discussed the factorization in nonleptonic $B$ meson decays and
have presented the LCSR calculation of the  $B \to \pi \pi $ decay.
It has been shown that the leading contribution factorizes, while
the corrections beyond the factorization can be systematically approached.
In the $m \to \infty$ limit, our result agrees with the QCD factorization prediction, 
while at finite $m_b$ we have found $O(\alpha_s/m_b)$ effects which are numerically 
small, but accumulate to a sizable correction in the direct $CP$ asymmetry. Large 
charming-penguin contributions {\it per se} are not predicted by this model.  
                                                                                         
Numerically, the nonfactorizable corrections in $B \to \pi \pi $ are not
large, but
the situation in the nonleptonic color-suppressed decays seems to be somewhat different.
Recent measurements of the color-suppressed $\overline{B}^0 \to D^{(\ast)0} \pi^0$ decays, $B \to J/\psi K$
and particularly
$B \to \chi_{cJ} K$ decays show large discrepancy with the naive factorization prediction and provide clear
evidence for large nonfactorizable contributions. The $B \to \chi_{c0} K$ and $B \to \chi_{c2} K$ decays
 are particularly interesting
because their  branching ratios predicted in
the naive factorization are exactly zero, and the measurements yield
branching ratios $\sim {\cal O}(10 ^{-4})$ which are therefore comparable with $BR(B \to J/\psi K)$.
This should not come out as a surprise, because
the LCSR calculation
shows the existence of large
nonfactorizable corrections in $B \to J/\psi K$ of the order of $70 \%$ 
\cite{Melic}. The corresponding large corrections 
are also expected for $B \to \chi_{cJ} K$ decays. 
Unfortunately, these large corrections still appear to be insufficient to reproduce the data, because 
the discrepancy between the the theory and the experiment is much larger (in the case of 
$B \to J/\psi K$, the data and the LCSR improved prediction still differ by a factor of two) and 
at the moment it is not clear how these discrepancies could be reduced. 
                                                                                         
\section*{Acknowledgments}
I would like to thank the organizers for the invitation to participate
at this very pleasant and  stimulating conference and for their
financial support.
This work is also supported by the Ministry of Science and Technology of the 
Republic of Croatia under the contract 0098002.


%
\label{MelicEnd}
                                                                                         
\end{document}